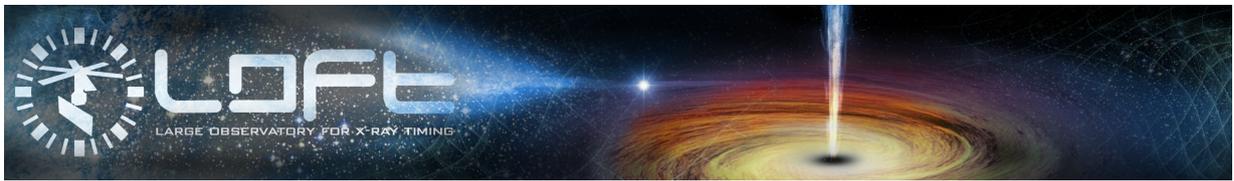

# Probing stellar winds and accretion physics in high-mass X-ray binaries and ultra-luminous X-ray sources with *LOFT*

## White Paper in Support of the Mission Concept of the Large Observatory for X-ray Timing

### Authors

M. Orlandini[1], V. Doroshenko[2], L. Zampieri[3], E. Bozzo[4], A. Baykal[5], P. Blay[6],
M. Chernyakova[7], R. Corbet[8], A. D'Aì, T. Enoto[9,16], C. Ferrigno[4], M. Finger[10],
D. Klochkov[2], I. Kreykenbohm[11,12], S.C. Inam[13], P. Jenke[14], J.-C. Leyder[15], N. Masetti[1],
A. Manousakis[16], T. Mihara[17], B. Paul[18], K. Postnov[19], P. Reig[20], P. Romano[21], A. Santangelo[2],
N. Shakura[19], R. Staubert[2], J.M. Torrejón[22], R. Walter[4], J. Wilms[11,12], C. Wilson-Hodge[23]

[1] INAF/IASF-Bologna, via Gobetti 101, I-40129 Bologna, Italy
[2] Institut für Astronomie und Astrophysik, Sand 1, 72076 Tübingen, Germany
[3] INAF-Astronomical Observatory of Padova, I-35122 Padova, Italy
[4] ISDC, University of Geneva, Chemin d'Écogia 16, 1290 Versoix, Switzerland
[5] Physics Department, Middle East Technical University, Ankara 06531, Turkey
[6] Image Processing Laboratory University of Valencia PO BOX 22085, E-46071, Valencia, Spain
[7] Dublin Institute for advanced studies, 31 Fitzwilliam Place, Dublin 2, Ireland
[8] University of Maryland, Baltimore County, Baltimore, MD 21250, USA
[9] NASA Goddard Space Flight Center, Astrophysics Science Division, Code 662, Greenbelt, MD 20771, USA
[10] National Space Science and Technology Center, 320 Sparkman Drive, Huntsville, AL, USA
[11] Dr. Karl Remeis-Sternwarte Bamberg, Sternwartstrasse 7, 96049 Bamberg, Germany
[12] Erlangen Centre for Astroparticle Physics, Erwin-Rommel-Str. 1, 91058 Erlangen, Germany
[13] Department of Electrical and Electronics Engineering, Baskent University, Ankara 06810, Turkey
[14] University of Alabama in Huntsville, 301 Sparkman Drive, Huntsville, Alabama, USA
[15] Research Fellow within the Science Operations Department, European Space Agency (ESA), European Space Astronomy Centre (ESAC), E-28691 Villanueva de la Cañada, Madrid, Spain
[16] Centrum Astronomiczne im. M. Kopernika, Bartycka 18, PL-00716 Warszawa, Poland
[17] MAXI team, RIKEN, 2-1 Hirosawa, Wako, Saitama, 351-0198, Japan
[18] Raman Research Institute, Bangalore 560080, India
[19] Moscow M.V. Lomonosov State University, Sternberg Astronomical Institute, 119992 Moscow, Russia
[20] IESL, Foundation for Research and Technology-Hellas, GR-71110 Heraklion, Crete, Greece
[21] INAF, Istituto di Astrofisica Spaziale e Fisica Cosmica – Palermo, via U. La Malfa 153, 90146 Palermo, Italy
[22] Instituto Universitario de Física Aplicada a las Ciencias y las Tecnologías, Universidad de Alicante, E03080 Alicante, Spain
[23] Astrophysics Office, ZP 12, NASA Marshall Space Flight Center, Huntsville, AL 35812, USA

## Preamble

The Large Observatory for X-ray Timing, *LOFT*, is designed to perform fast X-ray timing and spectroscopy with uniquely large throughput (Feroci et al., 2014). *LOFT* focuses on two fundamental questions of ESA's Cosmic Vision Theme "Matter under extreme conditions": what is the equation of state of ultra-dense matter in neutron stars? Does matter orbiting close to the event horizon follow the predictions of general relativity? These goals are elaborated in the mission Yellow Book (`http://sci.esa.int/loft/53447-loft-yellow-book/`) describing the *LOFT* mission as proposed in M3, which closely resembles the *LOFT* mission now being proposed for M4.

The extensive assessment study of *LOFT* as ESA's M3 mission candidate demonstrates the high level of maturity and the technical feasibility of the mission, as well as the scientific importance of its unique core science goals. For this reason, the *LOFT* development has been continued, aiming at the new M4 launch opportunity, for which the M3 science goals have been confirmed. The unprecedentedly large effective area, large grasp, and spectroscopic capabilities of *LOFT*'s instruments make the mission capable of state-of-the-art science not only for its core science case, but also for many other open questions in astrophysics.

*LOFT*'s primary instrument is the Large Area Detector (LAD), a $8.5\,\mathrm{m}^2$ instrument operating in the $2$–$30\,\mathrm{keV}$ energy range, which will revolutionise studies of Galactic and extragalactic X-ray sources down to their fundamental time scales. The mission also features a Wide Field Monitor (WFM), which in the $2$–$50\,\mathrm{keV}$ range simultaneously observes more than a third of the sky at any time, detecting objects down to mCrab fluxes and providing data with excellent timing and spectral resolution. Additionally, the mission is equipped with an on-board alert system for the detection and rapid broadcasting to the ground of celestial bright and fast outbursts of X-rays (particularly, Gamma-ray Bursts).

This paper is one of twelve White Papers that illustrate the unique potential of *LOFT* as an X-ray observatory in a variety of astrophysical fields in addition to the core science.

# 1 Summary

High mass X-ray binaries (HMXBs) constitute a sizeable fraction of the bright X-ray sources in the Milky Way and in other nearby galaxies. In these objects, the bulk of the X-ray emission is due to the accretion of material lost by a massive companion ($\gtrsim 10$ $M_\odot$) onto either a neutron star (NS) or a blackhole (BH). Depending mainly on the nature of the companion, accretion can be mediated by a disk or through a stellar wind.

Supergiant, hypergiant, and Wolf-Rayet stars have the densest, fastest, and highly structured winds. These winds can give rise to an extreme X-ray variability when accreted onto a compact object. Accretion from the milder winds of moderately massive red giants ($\gtrsim 1$ $M_\odot$) also produces a substantial X-ray variability. Our understanding of the relation between the X-ray variability and stellar wind properties has been limited so far by the lack of X-ray facilities featuring simultaneously a large collecting area and good energy and timing resolution. Accurate spectral and timing variability studies of wind-fed binaries require, at present, long integration times, leaving the study at typical dynamical to sub-dynamical time scales largely unexplored.

Young NSs (typically few $10^6$ yr) hosted in HMXBs are known to possess strong magnetic fields ($\gtrsim 10^{12}$ G) which channel the accreting material from distances as large as $\sim 10^3$-$10^4$ km down to their surfaces. This leads to the formation of extended accretion columns and complex pulse profiles with a remarkable energy dependence. Our understanding of the physics of this "magnetospheric accretion" has been hampered so far by the inability to reveal changes in the source spectral and timing properties within integration times shorter than a single source pulse period. Gaining access to such short integration times will permit investigation of the physics of the accretion columns, map the topology of the NS magnetic field, and ultimately reveal the state of matter under such extreme conditions.

The dynamics of plasma at the disk inner boundary of disk-fed NS HMXBs was also poorly investigated so far, as in these systems the disk is truncated at sizeable distances from the compact object ($\gtrsim 1000$ km) and the timing/spectral signatures produced by the moving plasma are weakly detectable. A significant improvement in the collecting areas of currently available X-ray facilities is needed to open-up studies in this field and transform current tentative detections in diagnostic tools.

Large accretion torques, that alternatively slow-down or spin-up the NS, have been measured on both wind and disk-fed HMXBs on time scales that range from years to decades. These torques and their reversals result from the yet unknown coupling between the NS magnetic field and the accreting material. Several models are still debated. Pursuing the long term monitoring of HMXBs is thus mandatory to secure continuity in these studies.

Ultra-luminous X-ray sources (ULXs) are also often found in young stellar environments (see, e.g., Liu et al., 2007; Grisé et al., 2008, 2011) and in at least two cases they have been directly associated with HMXBs (Liu et al., 2013; Motch et al., 2014). These sources are a challenge of modern X-ray astronomy, as the physical mechanisms producing their unusually high X-ray luminosity is still debated. This is partly due to the lack of high SNR X-ray timing/spectral data.

Thanks to the unique combination of collecting area and energy resolution in a wide energy band (2-50 keV), together with an unprecedented large field-of-view imaging more than 1/3 of the sky at once, the instruments on-board *LOFT* (Feroci et al., 2014) will dramatically open-up perspectives for research in all above mentioned fields. In particular:

- the LAD will grant for the first time access to accurate spectral and timing variability studies on dynamical and sub-dynamical time scales for wind-fed and disk-fed HMXBs. This will overcome the long standing difficulty in disentangling physical processes occurring in stellar winds, accretion columns, and the strongly magnetized environments of young pulsars. LAD observations will map the magnetic field topology of young NSs and investigate their microsecond variability. Detailed orbital monitoring observations will be possible for a large number of $\gamma$-ray-loud and $\gamma$-ray-quite HMXBs, as the LAD can measure changes in their emission properties within integration times as short as 1-10% compared to those required with the currently and past X-ray facilities.

- *Understanding the timing properties and searching for periodicities in bright ULXs.* The LAD will be able to perform a detailed timing analysis of a limited but non-negligible group of bright ($\geq$ a few tens of mCrab) sources. Linking timing properties to other source properties (e.g. spectral/flux variability) or to each other (high vs. low frequency QPOs) is of primary scientific importance to understand how they can be produced in a non-standard accretion regime, a crucial step to use them for estimating BH masses.

- the WFM will promptly detect fast flares and the onset of outburst from HMXBs, as well as discover new transient sources in this class and nearby ULXs ($\leq 4$ Mpc). It will be possible to monitor continuously a large fraction of bright accreting HMXBs, providing for them data with good spectral and timing resolution to study their long term variability and accretion torques. Synergies are envisaged between *LOFT* and the Square Kilometer Array to jointly monitor ULX radio/X-ray variability (Wolter et al. 2014).

## 2 Probing stellar winds in wind-fed binaries

### 2.1 Transient structures

Massive stars generate dense fast outflows that trigger star formation and drive the chemical enrichment and evolution of Galaxies (Kudritzki, 2002). The amount of mass lost through the emission of these winds has a large impact on the evolution of the star (see, e.g., Meszaros & Rees, 2014, for a recent review). In the past decades, observational evidence has been growing that winds of massive stars are populated by dense "clumps". The presence of these structures affect the mass loss rates derived from the study of stellar winds, thus leading to uncertainties in our understanding of their evolutionary paths (Puls et al., 2008).

*HMXBs were long considered an interesting possibility to probe clumpiness* (see, e.g., Sako et al., 2003, and references therein). A substantial fraction of these systems host a compact object orbiting an O-B supergiant and accreting material from the stellar wind (see, e.g., Chaty, 2013; Paul & Naik, 2011, for recent reviews). As the X-rays released by the accretion process can trace the mass inflow rate around the compact object, such an object provides a natural probe to measure *in situ* the physical properties of the massive star wind and clumps. The so-called "Supergiant Fast X-ray Transients" (SFXTs) showed so far the most convincing evidence for the accretion of large clumps onto the compact object (in't Zand, 2005; Walter & Zurita Heras, 2007). In X-rays, the role of clumps is two-fold. Clumps passing in front of the compact object cause (partial) obscuration of the X-ray source, and display the signatures of photoelectric absorption and photo-ionization (D'Aí et al., 2007; Rampy et al., 2009). In addition, clumps can lead to temporarily increased accretion and X-ray flares (Ducci et al., 2009; Oskinova et al., 2012). A number of hours-long flares displayed by the SFXTs could be convincingly associated to the accretion of dense clumps, the most striking case being that of a moderately bright flare from IGR J18410−0535 observed with *XMM-Newton* (Bozzo et al., 2011). With the Epic-pn camera, integration times of several hundreds to thousands seconds were needed to get a rough estimate of the clump properties and an average picture of the clump accretion process.

Similar observations performed with the LAD on-board *LOFT* will dramatically improve our present understanding of clumpy wind accretion and winds in massive stars in general. **By taking advantage of the unprecedentedly large effective area of this instrument, it will be possible to study spectral and intensity variability of the source on time scales as short as few to tens of seconds**. This will permit a study of the dynamics of the clump accretion process in detail and obtain more reliable estimates of the clump mass, radius, density, velocity, and photo-ionization state. The latter can be probed also by revealing changes in the centroid energy of the fluorescent iron line between 6.4-6.6 keV, commonly produced in wind-fed binary as a consequence of the X-ray irradiation of the wind and clump material (see, e.g., Torrejón et al., 2010, for a recent review).

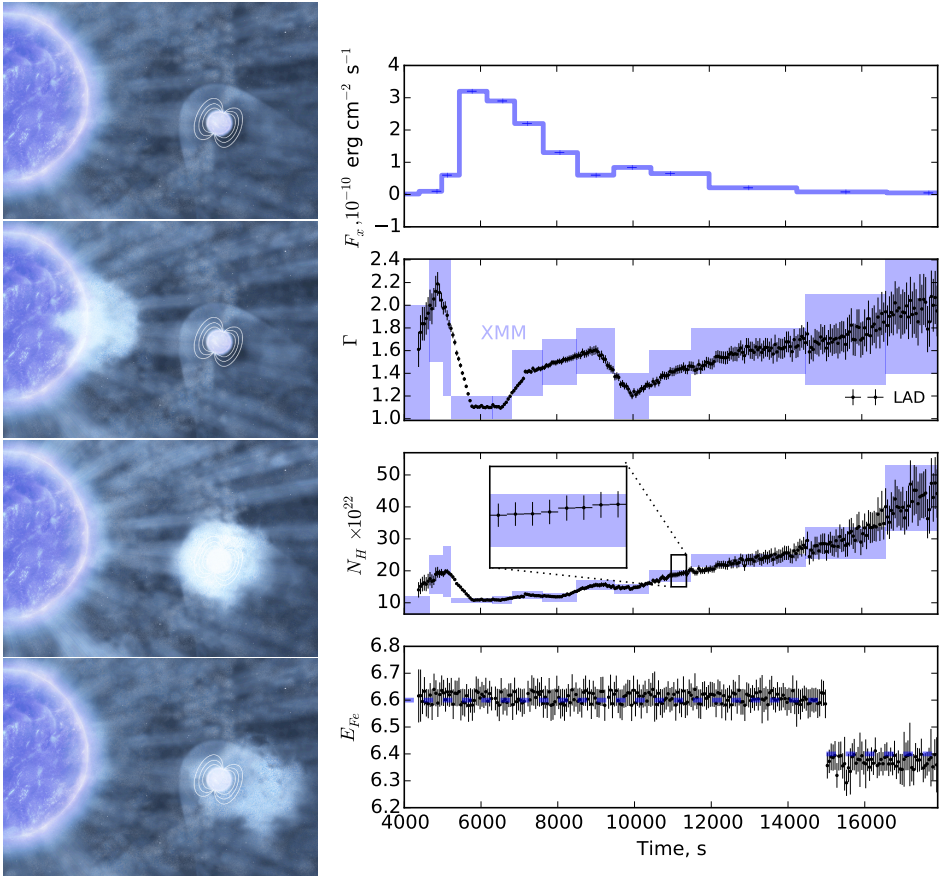

Figure 1: Left: Sketch of the accretion of a clump (credits: ESA). *Right*: Changes in the flux and spectral parameters during the flare recorded by *XMM-Newton* from IGR J18410−0535. Violet boxes represent the measurements obtained with *XMM-Newton*, while black points represent values obtained from the simulated LAD spectra with exposure times as short as 50 s (uncertainties are sometimes too small to be visible on the scale of the figure). The LAD sensitivity to changes in the centroid energy of the fluorescent iron line (6.4-6.6 keV) is addressed in the bottom panel.

Based on the spectral parameters measured in *XMM-Newton* observation of IGR J18410−0535, we simulated several LAD spectra over the course of the flare by using integration times as short as 50 s and interpolating the spectral properties of the source between the individually available *XMM-Newton* points. As is shown in Fig. 1, during similar events, the LAD is able to reveal spectral changes in the continuum as well as in the iron line energy with a better accuracy than current facilities by using integration times a factor of ∼10-100 shorter. Similar studies will be possible with other sub-classes of wind-fed X-ray binaries as well, testing winds and clumps parameters in different types of stars.

### 2.1.1   The case of symbiotic X-ray binaries

Bright wind-accreting X-ray binaries can also be found among low mass X-ray binaries (LMXBs), in particular in the so-called Symbiotic X-ray binaries (SyXBs). These are interacting systems in which the compact object moves in a wide orbit (periods months to years) around a red giant and accretes material from its slow and dense wind (Masetti et al., 2006; Nespoli et al., 2010; Hynes et al., 2014; Bahramian et al., 2014). Observationally, SyXBs are characterized by a highly variable X-ray emission, displaying bright flares similar to those of other wind-fed X-ray binaries (Masetti et al., 2007). X-ray pulsations are unusually slow, but demonstrate unambiguously the presence of accreting NSs in these sources (spin periods range from hundreds to ∼18400 s; Corbet et al., 2008). About a dozen SyXBs are known to date, and the majority of these systems are characterized by average X-ray fluxes of $10^{-11}$-$10^{-10}$ erg cm$^{-2}$ s$^{-1}$.

**LAD observations can thus be carried out as in the case of HMXBs to probe the properties of the poorly known winds from red giants** (which are dust rather than radiatively accelerated) and probe the physics of

Figure 2: The quasi-spherical accretion model is particularly well suited to wind accreting system in which the mass accretion rate is low (figure from Enoto et al., 2014). In this model a hot quasi-static shell forms above the NS magnetosphere and plasma entry through the magnetic field lines is regulated by inefficient radiative plasma cooling or Compton cooling (Shakura et al., 2013). The application of this model to the case of 3A 1954+319, a SyXB prototype, would solve the problem of the peculiarly large value of the magnetic field above the magnetar range ($10^{16}$ G), required by the standard wind accretion theory (see Enoto et al., 2014, and references therein). We note that the quasi-spherical accretion model is also applicable to other classes of wind-fed sources (e.g., the SFXTs), and thus a better understanding of this accretion regime is of wide interest (Shakura et al., 2014, 2013).

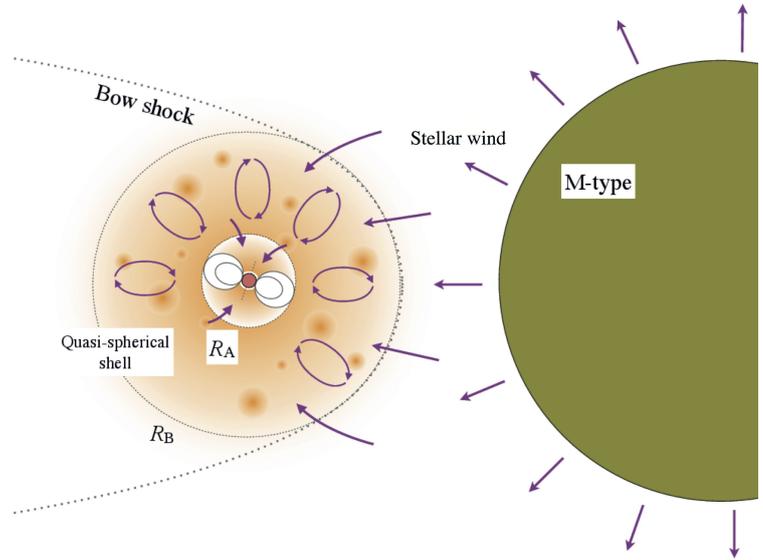

wind accretion from slow flows at low mass loss rates ($<10^{-10}$ $M_\odot$ yr$^{-1}$ compared to the typical $10^{-5}$-$10^{-6}$ $M_\odot$ yr$^{-1}$ observed in supergiant stars). The latter are particularly interesting to test the quasi-spherical accretion model onto magnetized NSs (see Fig. 2; Davies et al., 1979; Ikhsanov, 2007; Shakura et al., 2012), and the X-ray variability model via red giant wind accretion suggested by Filippova et al. (2014). The continuous sky monitoring performed with the WFM will increase the number of SyXBs by, e.g., discovering long period pulsations (see Sect. 4).

### 2.1.2 The case of wind colliding binaries and Eta-Carinae

Another interesting class of binaries for which *LOFT* observations could help probing the properties of star winds is that of the colliding "wind binaries" (see, e.g., Stevens et al., 1992). Among these, we consider below specifically the case of $\eta$ Car, a nearby massive binary system (distance of 2.3 kpc) hosting a luminous blue variable and a Wolf-Rayet or a late-type nitrogen-rich O star, in a 5.5 yrs orbit. The source is highly absorbed, embedded in a thick cocoon due to episodes of massive ejections (see, e.g., Davidson & Humphreys, 1997). The X-ray emission above 2 keV is dominated by a point-like optically thin thermal component with a temperature of ~4 keV which is most probably due to the collision between the two stellar winds. Regular weekly monitoring by the *RXTE* PCA has revealed a periodic variability linked to the orbital motion with a maximum at periastron lasting for a few months, followed by a strong 3-month long suppression of soft X-ray emission. A phase known as deep minimum sees the almost complete suppression of the soft X-ray component, possibly due to occultation by the stellar photosphere (see Corcoran, 2005, and references therein). Shorter term variability is also present and its origin might be linked to wind clumpiness (Moffat & Corcoran, 2009). An excess power-law emission has been reported above ~ 10 keV (Viotti et al., 2004; Sekiguchi et al., 2009; Hamaguchi et al., 2014; Leyder et al., 2010) with an index of 1-1.8 and a 10-50 keV flux of 1-2×$10^{-11}$ erg/s/cm$^{-2}$. This component is thought to extend up to the GeV range, where it can be matched to the FERMI detection (Farnier et al., 2011); it might be interpreted as a signature of particle acceleration at the shock between the winds, but its origin is debated.

*LOFT* will be able to perform both monitoring and detailed X-ray spectral analysis owing to the large effective area of the LAD. We have simulated a 3 ks long snapshot assuming the spectral parameters determined with Suzaku (Hamaguchi et al., 2014) for ten orbital phases with a constant power-law emission along the orbit but variable soft X-ray emission. *We have verified that the power-law can be firmly detected even at the soft X-ray maximum, and determining the index more precisely than Suzaku requires only <10% of its observing time*. If the LAD instrument observes $\eta$ Car every few weeks during 3 (satellite) revolutions, it will help understand the mysterious nature of the hard X-ray emission, and it will allow to search for its variability. The hard X-ray emission can be used to constrain the characteristics of the particle acceleration in the colliding wind region, such as the particle density and the magnetic field. **These monitoring observations could also be useful to quantify the mass loss rates of the stellar winds and their clumped nature,** by a detailed study of the relevant spectral/variability parameters (see also Sect. 2). We remark that this cannot be easily done at other wavelengths as the radiation is heavily absorbed.

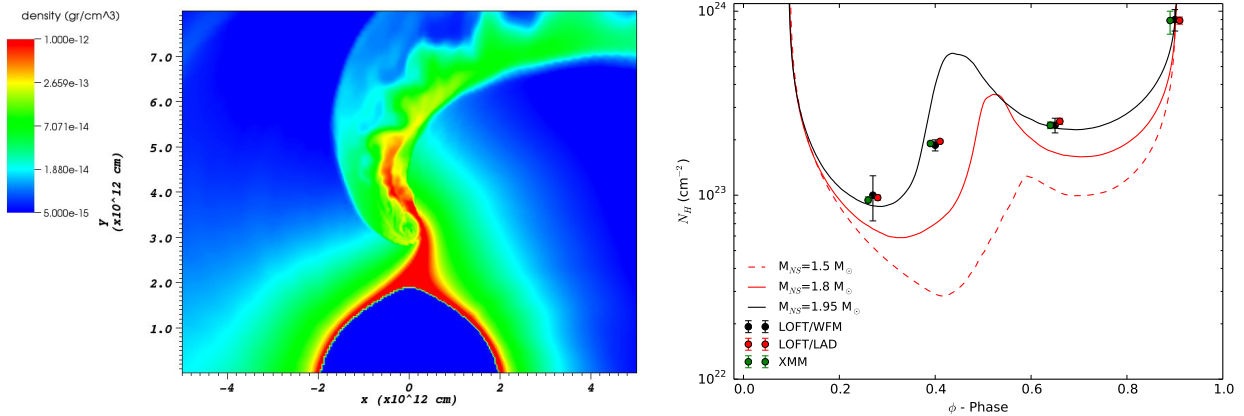

**Figure 3:** *Left:* An example of hydrodynamical simulations from Manousakis et al. (2012). The wind of the primary is strongly perturbed by the gravitational field of the NS. A dense stream of matter trails the NS and absorbs its X-ray emission. *Right:* Absorption column density as a function of the orbital phase and NS mass, as measured in the case of IGR J17252−3616 (average X-ray flux of $F_x = 5.5 \times 10^{-11}$ erg cm$^{-2}$ s$^{-1}$ in the 3-30 keV energy range; Manousakis & Walter, 2011). The green points are the original *XMM-Newton* measurements (≳10 ks exposure for each point), while red points represents the simulated LAD observations by using only 1 ks of exposure. We also show the results obtained from the simulated WFM spectra (black points) for a brighter source with a flux similar to Vela X-1 ($F_x = 5.5 \times 10^{-9}$ erg cm$^{-2}$ s$^{-1}$). The WFM achieves, for such fluxes, a good accuracy in the measurement of the spectral parameters already in ≲20 ks.

## 2.2 Persistent structures

Many wind-fed HMXBs display prominent variations of their average X-ray spectral properties at different orbital phases. This variability occurs on a much longer time scale with respect to that discussed in Sect. 2.1 and provide different important insights in persistent structures that form around the compact object when the stellar wind is focused by its gravitational field (Blondin et al., 1991).

Recently Manousakis & Walter (2011) investigated the variations of the X-ray emission properties from the wind-accreting supergiant X-ray binary IGR J17252−3616. By using *XMM-Newton* observations, these authors found that the absorbing column density has a peculiar dependence on the orbital phase (see Fig. 3). Manousakis et al. (2012) used hydrodynamic simulations to show that the increase of the absorption column density is most likely caused by an accretion wake that form around the NS. The phase profile of the absorption column density could be well reproduced in the simulation, suggesting that the mass of the compact object plays a key role. *If extended to other wind accreting systems, these studies could provide a new independent way to measure NS masses.* A first attempt to extend the applicability of this study to the supergiant X-ray binary prototype Vela X-1 was presented by Doroshenko et al. (2013). However, for most wind-fed binaries, a sufficiently accurate monitoring of the absorption column density (as well as other relevant spectral parameters) is not yet available as such observations requires relatively long exposure times with the current X-ray facilities.

**Figure 4:** An "off-state" observed from the source GX 301-2 with the *RXTE*/PCA (Göğüş et al., 2011) and a simulation of *LOFT* capabilities to study similar events. The grey area represents the source lightcurve as observed by the PCA. The blue show the accuracy and integration time of the source spectral slope measured by the PCA ($\Gamma$ is the power-law spectral index). The black points show how the LAD could be used to follow the spectral variations in greater detail by taking advantage of its large collecting area and good spectral resolution (we simulated LAD spectra in 50 s integration times and interpolated the photon index values among the PCA measurements). The red points and the corresponding error-bars represent the WFM lightcurve of the source simulated with a time-bin of 685 s (right scale). Note that despite the significant drop in flux, the source can still be detected by the WFM during the off-state.

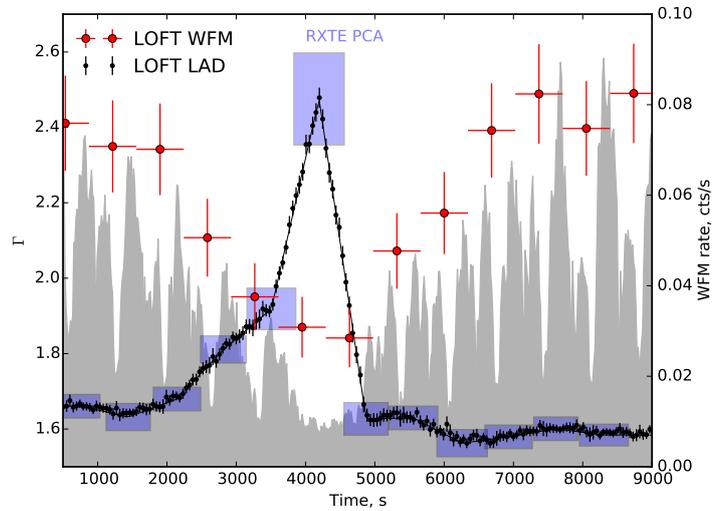

The large effective area and good broadband coverage of the LAD make it perfectly suited to carry out these investigations, as the required measurement accuracy can be obtained with much shorter time scales compared to, e.g., *XMM-Newton* or MAXI (see, e.g., Islam & Paul, 2014, for a recent study on the orbital phase resolved spectroscopy of GX 301-2). In Fig. 3 we present a comparison of the results obtained for IGR J17252−3616 with the Epic-pn on-board *XMM-Newton* and a simulation of the same observations with the LAD. **The LAD can achieve the same accuracy in the spectral parameters of the source by using ≲1/10 of the Epic-pn exposure time** (note that for the majority of the systems of interest $N_H \gtrsim 10^{22}$ cm$^{-2}$). For brighter sources like Vela X-1, WFM data would also be able to provide the required long-term spectroscopic information, without the need of dedicated LAD campaigns (see Fig. 3).

## 2.3 Probing the origin of off-states in wind-fed binaries

An intriguing and poorly understood phenomenon in wind-fed binaries is that of the so-called "off-states" (Kreykenbohm et al., 2008; Göğüş et al., 2011; Doroshenko et al., 2012). These are rapid (few seconds) and sporadic drops of the source X-ray flux, lasting typically several pulsation cycles, which origin is currently unknown. The study of such events has so far been hampered by their short duration, which requires long monitoring observations to be discovered in different sources, and the inability to obtain systematically accurate spectral and timing information during the ingress and egress phases (Doroshenko et al., 2011). Different spectral changes would be expected if the off-states are due to: (i) accretion from regions characterized by a rarefied stellar wind (i.e., the intra-clump medium; Kreykenbohm et al., 2008; Oskinova et al., 2012); (ii) switches in magnetospheric accretion modes (Bozzo et al., 2008; Doroshenko et al., 2011); (iii) switches between the Compton and radiatively inefficient cooling modes in the settling accretion regime (Shakura et al., 2013); (iv) hydrodynamic instabilities close to the compact object magnetosphere (Manousakis & Walter, 2014). Changes in the source power spectrum during the ingress and egress phases could also help revealing switches between different accretion modes or dynamical configuration of the accretion flow.

**The large effective area of the LAD will provide for the first time detailed spectral and timing informa-**

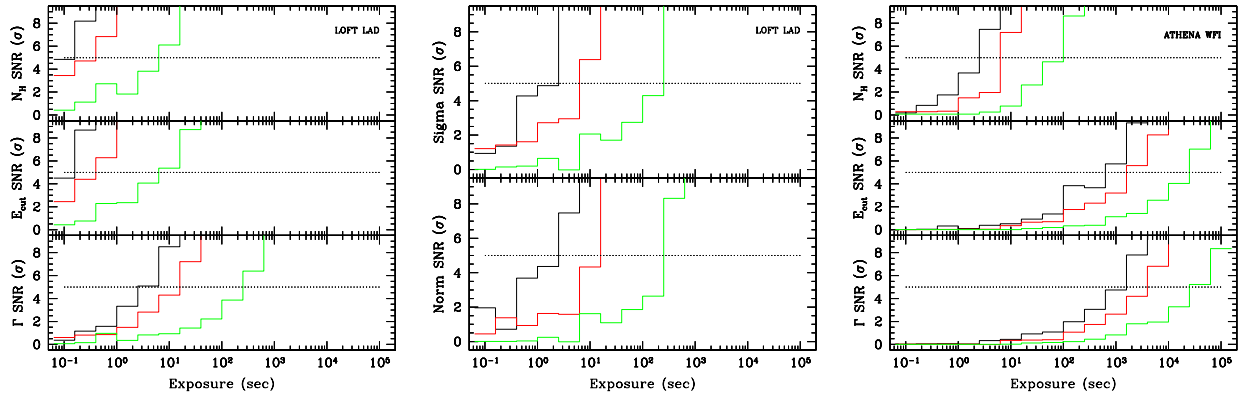

**Figure 5:** Example of the LAD capability to reconstruct the spectral shape of an accreting X-ray pulsar in very short integration times. Simulations were performed using the results of a BeppoSAX observation of Vela X-1 (Orlandini et al., 1998). Each curve gives the significance of a spectral parameter (defined as the ratio between the best fit parameter and its uncertainty) as a function of the exposure time. The curves are color coded according the the assumed flux (black: 500 mCrab, red: 150 mCrab, green: 15 mCrab). *Middle:* Same as before but showing the LAD capabilities in the case of 4U 0115+634. For this source we also show the significance of the CRSF detection at 12.8 keV using the spectral model proposed by (Santangelo et al., 1999). "Norm" is the CRSF normalization and "Sigma" its width. We do not plot the significance of the CRSF centroid energy, as it is always determined at a very high significance. The LAD can indeed measure the CRSF centroid energy at a significance level of 100, 20, and $3\sigma$ for fluxes of 500, 150, and 15 mCrab, respectively in only 0.1 s. *Right:* Same as the left panel, but for Athena/WFI. This show that similar studies are only possible if the LAD large collecting area is available. Other currently planned X-ray instruments would need too long exposure times to achieve accurate spectral measurements.

**tion within integration times shorter than the off-state ingresses and egresses**. This is illustrated in Fig. 4 by using the example of the off-states in GX 301-2 discovered with the *RXTE*/PCA (Göğüş et al., 2011). In the figure we also show that off-states can be easily detected with the WFM, by taking advantage of its good sensitivity and the availability of long term monitoring observations of many wind-fed binaries during the course of the mission (see also Sect. 4).

## 3 Probing the physics of magnetospheric accretion

### 3.1 Pulse phase resolved spectroscopy and cyclotron features

Close to the magnetic poles of a young NS, X-ray photons are forced to interact with the magnetic field before being able to escape in the direction of the observer. As the quantization of the electron motion in the direction perpendicular to the magnetic field has the consequence that the scattering cross sections of X-rays are anisotropic and energy dependent, X-ray photons with frequencies close to the gyro-magnetic (Larmor) frequency are scattered out of the line of sight, giving rise to "cyclotron resonant scattering features" (CRSFs). These features are often observed in the X-ray spectra of highly magnetized NSs and provide a unique tool to measure the compact object magnetic field strength (Canuto & Ventura, 1977). Because the CRSF parameters strongly depend on the viewing angle, a spectral analysis as a function of the pulse phase (pulse phase spectroscopy, PPS) can be used to reconstruct the topology of the NS magnetic field and probe the condition of matter close to the emitting regions. Such studies have been hampered so far by the fact that PPS could only be performed by averaging the source spectrum at a given phase over many individual pulses. This provides information only on the averaged large scale magnetic field configuration and spectral formation mechanisms, limiting the development of theoretical models (see, e.g., Schönherr et al., 2014, for a recent review). These topics thus remain highly debated (Staubert et al., 2007; Schönherr et al., 2007; Nishimura, 2011; Becker et al., 2012; Poutanen et al., 2013).

Figure 6: Cyclotron line energy $E_{cyc}$ as a function of luminosity in V 0332+53. Measurements at high luminosities (black open circles) are from Tsygankov et al. (2010). LAD measurements are simulated by carrying out PPS on one 20 ks-long observation at lower fluxes (red filled circles) and another at higher fluxes (green squares; error bars are smaller than the marker size in this case). LAD observations at higher and lower fluxes are (clearly) also doable but not shown here. A transition between different $E_{cyc}$-flux dependencies can be probed for the first time on a wide range of luminosities.

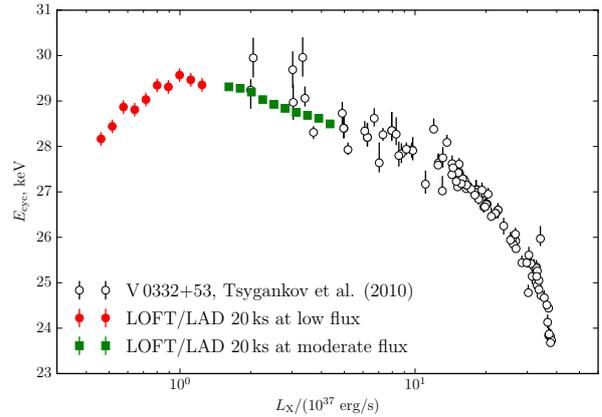

**The high throughput of the LAD will be able for the first time to accurately constrain spectral properties of accreting X-ray pulsars by using integration times much shorter than a single NS pulse**. This is illustrated in Fig. 5 for two X-ray pulsars prototypes: Vela X-1 and 4U 0115+634. *It is worth mentioning that, given the optimal response and effective area of the LAD at 6-8 keV, PPS analyses will not be limited to CRSFs, but naturally encompass any spectral feature, e.g., the fluorescence iron-K emission line (simulations show that the typical iron line observed in Vela X-1 can be constrained by the LAD with an integration time ≲10 sec).*

The LAD will also be able to perform PPS analyses at low luminosities, a regime never investigated so far due to the lack of simultaneous broad-band energy coverage, energy resolution, and collecting area of past and present X-ray facilities. Especially intriguing is the variation of the cyclotron line energy with luminosity observed in several sources (see, e.g., Klochkov et al., 2011b). Such a variation most probably reflects a vertical displacement of the emitting region in the inhomogeneous magnetic field of the NS, and could be investigated so far only at luminosities ≳$10^{37}$ erg s$^{-1}$ with relatively long pointed *RXTE* and *INTEGRAL* observations (see, e.g., Klochkov et al., 2011a, and references therein). At luminosities below ~$10^{37}$ erg s$^{-1}$, the polar emission regions of the NS would be expected to switch to a different configuration leading to a different dependence of the cyclotron line energy $E_{cyc}$ on flux (Becker et al., 2012). **Simulations in Fig. 6 show that the LAD will be able to measure spectral parameters of the continuum of accreting pulsars at high significance (≳5 $\sigma$) in less than a few 100 s even at luminosities as low as few tens of mCrab**.

## 3.2 Inside the NS magnetosphere: microsecond variability in HMXBs

In wind-accreting NSs, the plasma is believed to penetrate the compact object magnetic barrier through the Rayleigh-Taylor instability, in the form of accreting blobs (Arons & Lea, 1976; Shakura et al., 2012). These blobs, after a short radial infall, are channeled by the magnetic field lines toward the NS polar caps, where they release their kinetic energy into X-ray radiation. A certain "granularity" in the observed X-ray emission is thus expected (Orlandini & Morfill, 1992). It was shown that the energy release of each shot occurs on microsecond timescales (Orlandini & Boldt, 1993). The passage of the microsecond pulses through the NS magnetosphere will tend to broaden them, but photons moving in a direction parallel to that of the magnetic field and energies much less than the cyclotron energy would emerge unscattered (because their scattering coefficient are drastically reduced; Herold, 1979). From the typical cyclotron resonance energies observed in XRPs, it is expected that for $E ≲ 5$ keV, $\mu$s pulses from the surface of NSs should be detectable relatively undistorted.

**The technique we are planning to use to detect such $\mu$s variability is multiple detector coincidence. The LAD is therefore the best suited instrument for this kind of studies, given its time resolution and high number of independent detection modules,** $N_d$. The presence of events at $\mu$s time scales can be probed by

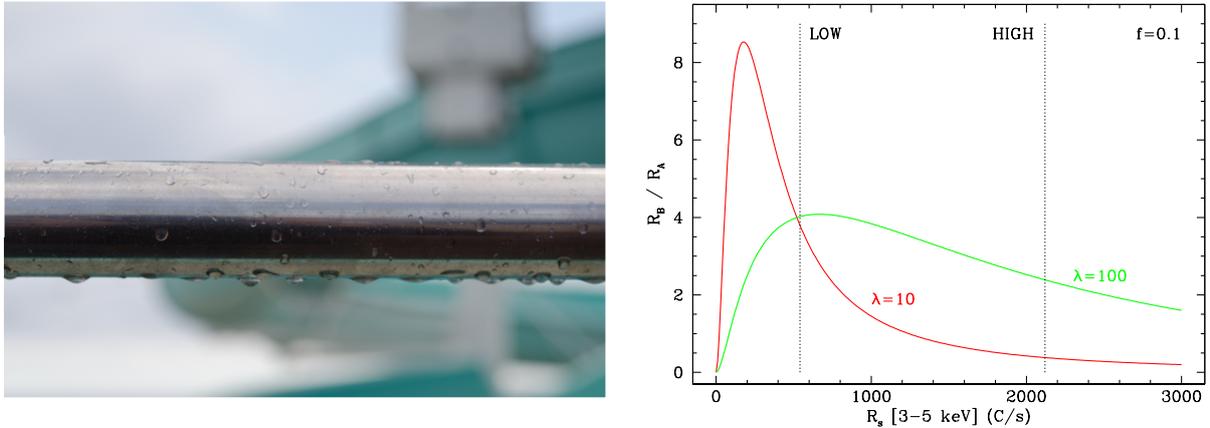

Figure 7: *Left*: The way in which matter can penetrates the magnetic boundary resembles the way drops of rain drip from a handrail. The drops of matter enter the magnetosphere through the Rayleigh-Taylor instability, "raining" onto the magnetic poles of the NS and generating $\mu$s variability. *Right*: *LOFT* unique capabilities in detecting microsecond bursting activity from wind-accreting NSs. The ratio between the detection rate $R_B$ of true micro-bursts and the rate $R_A$ of spurious bursts, is plotted as a function of the source count rate $R_s$, for the case of a moderate granularity (10%) in the accretion stream. The two vertical lines mark the count rates for two Vela X-1 luminosity states (low: 15 mCrab, soft spectrum; high: 45 mCrab, hard spectrum). $\lambda$ is the rate of formation of the blobs at the magnetospheric limit (Orlandini & Morfill, 1992). This would correspond to the number of rain drops formed in the handrail per unit time. $f$ is the granularity component in the accretion flow, here assumed to be 10%.

building a time interval histogram with 5 $\mu$s time resolution. By performing multiple detector coincidence, it is possible to discriminate between events that occur within 5 $\mu$s in different detectors. If all X-rays come to the detectors uncorrelated and at a constant rate, the histogram is expected to follow an exponential decay. Any deviation from this trend indicates coherence or correlation among the detected events. In particular, microsecond X-ray bursts could be identified in the histogram as excesses in some bins. Following Orlandini & Boldt (1993), we computed in Fig. 7 the rate $R_B$ of true bursts from coincidence timing among independent detectors, and the rate $R_A$ of *accidental* counts within 5 $\mu$s (e.g., coming from background fluctuations) expected for a LAD observation of Vela X-1 in the low and high emission state (Orlandini et al., 1998). According to the simulation, the number of detected events (bursts plus accidentals) would be greater than the expected number of accidental events even for a relatively small granularity of 10% in both the low and high emission states. For example, for the low state the rate of multiple coincidence detection of true $\mu$s bursts is $R_B$=14.1 c/s, to be compared to the expected rate of observing accidental coincidences $R_A$=4.3 c/s (note that this is the rate of true bursts from multiple detector coincidence, not to be confused with the rate of formation of the blobs at the magnetospheric limit, $\lambda$, which is in the range 10-100 blobs/s; it is the squashing of these blobs that gives origin to the $\mu$s variability, as detailed in Orlandini & Morfill, 1992). It is important to remark that the SNR for the detection of microsecond variability scales with the square root of $\binom{N_d}{2}$. The *RXTE*/PCA had $N_d = 5$, therefore the number of couples of detection modules among which to perform coincidence was $\binom{N_d}{2} = 10$. *For the LAD, $N_d$=100, and thus the S/N improvement over the PCA is $\gtrsim$70.* **No other flown or planned X-ray instrument is thus capable to efficiently perform such studies but the LAD.**

Variability on similar time scales is also expected in HMXBs due to the so called "photon bubble oscillations". The latter were found to develop below the radiation dominated shock that terminates the free-fall motion of the accreting matter in the accretion column of X-ray pulsars with luminosities $\gtrsim 10^{37}$ erg s$^{-1}$ (Klein et al., 1996). *If convincingly detected, photon bubble oscillations can provide insights on the structure of the accretion column near the NS surface and potentially an independent measurement of the compact object magnetic field* (Klein et al., 1996). Some evidence for the presence of variability induced by photon bubble oscillations was found in the HMXB Cen X-3, as two quasi-periodic-oscillations (QPOs) at 330 and 760 Hz have been reported in its power spectrum (Klein et al., 1996). Such features could not be convincingly confirmed with the 5 PCUs of the *RXTE*/PCA, as dead time effects lowered the counting noise level of the instrument below the value for expected for Poisson statistics and a counting noise model needed to be included in the fit to the power spectra. In this case, the large number of detectors in the LAD provides again a unique advantage.

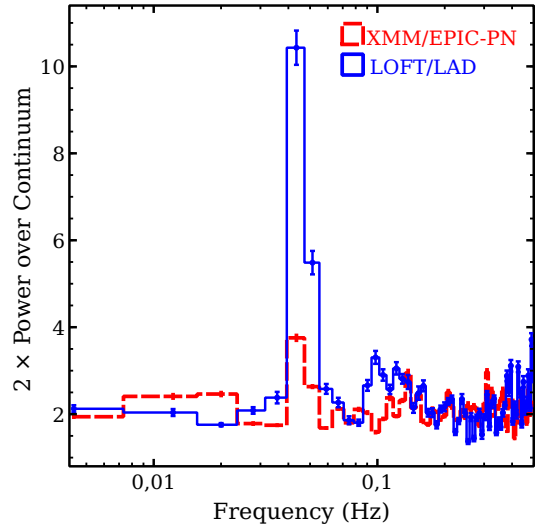

Figure 8: Power spectrum density (PSD) of the X-ray emission from SAX J2103.5+4545 recorded by the Epic-pn (red dashed line) in 2004, when the transient QPO at 44 mHz was observed. The exposure time is 4.2ks and the detection significance is 4 $\sigma$. The simulated LAD PSD of the same observation is shown with a blue solid line. A prominent feature is observed at >7 $\sigma$. Both PSDs were rebinned by a factor of 8, multiplied by 2, and divided by the continuum fit consisting of a broken–power law model with the power indices 0.34 and 2.14. The QPO was simulated in the 1 s-binned LAD lightcurve assuming a Lorentzian with a FWHM of 6.1mHz.

### 3.3   Probing the state of matter at the inner disk boundary in disk-fed HMXBs

QPOs at mHz frequencies in NS HMXBs have been observed in a number of systems, but their origin is still highly debated. They are tentatively interpreted as being related to the motion of material close to the inner boundary of the accretion disk (see, e.g., Finger, 1998; Bozzo et al., 2009, and references therein). As highly magnetized NSs are able to truncate the disk at distances of $10^3$-$10^4$ km from their surface, these features are weakly detectable (see Sect. 1), but hold the potential to provide measurements of the NS magnetospheric radius and an independent estimate of the NS magnetic field strength (see, e.g., Ghosh et al., 1977; Wang, 1995; Bozzo et al., 2009, and references therein). **Observing mHz QPOs with a sufficiently high SNR, following also their evolution with the source X-ray flux, requires particularly large collecting areas and thus the LAD is a perfectly suited instrument to conduct these studies**.

To illustrate the LAD capabilities in this field, we show in Fig. 8 the case of a weak mHz QPO detected from the Be X-ray binary SAX J2103.5+4545 (İnam et al., 2004). The comparison between the detection significance of the LAD, as compared to the Epic-pn, is striking. The LAD will be able to detect similar features in much fainter objects and follow their evolution across a wide range of X-ray luminosities. Our simulations showed that the QPO from SAX J2103.5+4545 would have been observed at high significance level (6 $\sigma$) by the LAD in only 500 s of exposure.

## 4   Monitoring HMXBs

The WFM is a powerful instrument, combining for the first time imaging capabilities, an unprecedentedly large field-of-view (FoV), and a good energy/timing resolution. This instrument will be capable of monitoring continuously hundreds of X-ray sources a day, providing for each of them high quality data and enabling innovative science even without the need of pointed LAD observations. As many HMXBs are transient, the discovery of new sources or new outbursts from previously known objects in this class will greatly benefit from the capabilities of the WFM. **In Fig. 9, we show that a large fraction of all kinds of sources described in the previous sections can be efficiently monitored during each single pointing of the WFM**, e.g. while *LOFT* is observing toward the Galactic center. As it is not possible here to exhaustively describe all the interesting scientific investigations enabled by the WFM, a number of examples are described in the next sections.

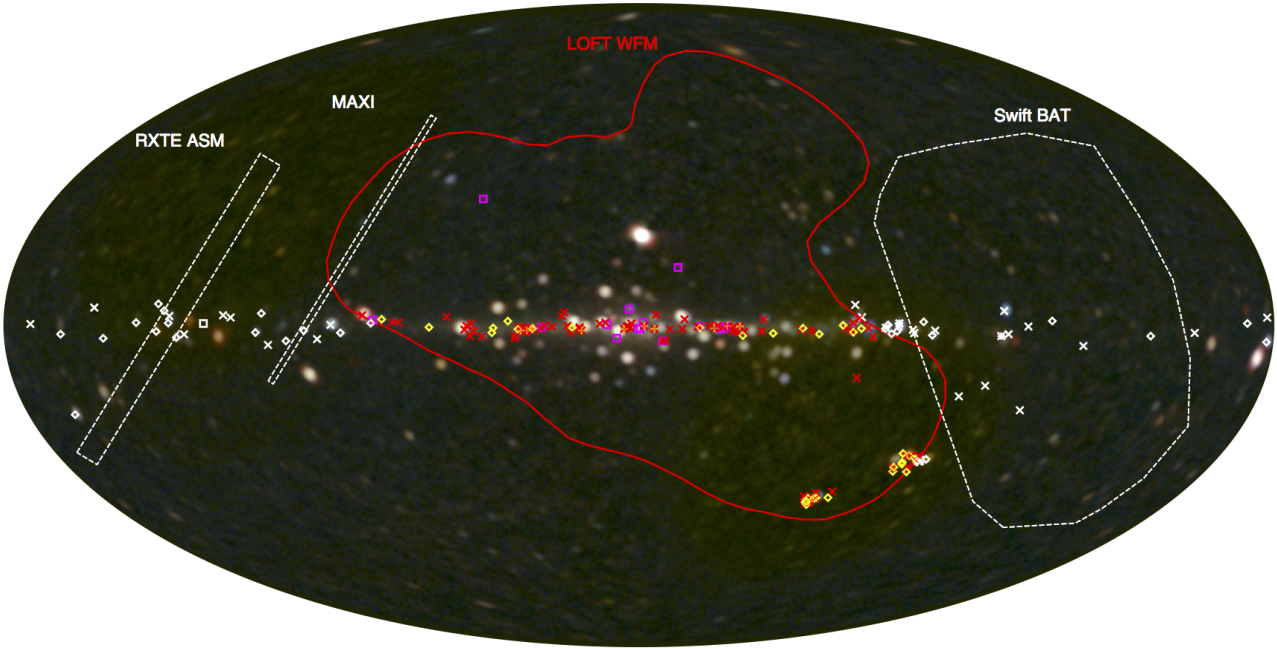

**Figure 9:** Comparison of the WFM FoV and the most relevant existing facilities (background map courtesy of RIKEN, JAXA, MAXI team). The WFM will have the largest FoV and monitor simultaneously a large fraction of all currently known HMXBs. In the map we represent known Be X-ray binary systems with yellow diamonds, supergiant X-ray binaries with red crosses, SFXTs with orange crosses, and SyXBs with magenta boxes. Grayed-out sources are those falling outside the WFM FoV during a single pointing toward the Galactic Center.

## 4.1 Accretion torques in disk-fed and wind fed X-ray pulsars

Over the past 14 years, observations with CGRO, *RXTE*, *INTEGRAL*, and *Swift* increased the number of known accreting X-ray pulsars from the 44 reported by Bildsten et al. (1997) to more than 120. Many of these are hosted in HMXBs, but others also in SyXBs and in intermediate systems between HMXBs and LMXBs. *The large FoV instruments onboard these missions provided the opportunity to continuously monitor spin period changes of these objects, even though not all of them were capable to provide spectral and fluxes information.* Particularly striking were the findings for GX 1+4 and 4U 1626-67 : after about 15 yr (for GX 1+4) and 20 yr (for 4U 1626-67) of spin-up, both systems showed a torque reversal, which made them switch to a spin-down phase. Other systems, such as Cen X-3, Vela X-1, and Her X-1, often showed an alternation of spin-up and spin-down sometimes superimposed on a longer term trend of either spin-down or spin-up. *Spin period changes in all these systems are known to be associated with accretion torques, with torque reversals reflecting dramatic changes in the interaction between the NS magnetic field and the accretion disk* (see, e.g., Perna et al., 2006, and references therein). Different models have been proposed, but a widely accepted scenario to explain the different behaviour observed in disk and wind-fed systems is still lacking. Long-term monitoring of these sources is the key to understanding these phenomena and answering fundamental questions such as: (i) is the intensity of torques always proportional to the flux of the source and thus to the mass accretion rate? (ii) What triggers torque reversals? (iii) What is characterizing spin-up and spin-down states other than the frequency rate and how do these states evolve between transitions? Theoretical models give a relationship between the spin-up rate and luminosity, as a function of the NS mass, radius, and magnetic field (Bildsten et al., 1997). For systems with well known distances, simultaneous measurements of fluxes and spin-rates can provide constraints on the magnetic field (Coe et al., 2014; Klus et al., 2014). If the pulsar magnetic field is also known from cyclotron line measurements, then the relationship between the spin rate and the luminosity can be used to constrain the NS parameters, probing not only accretion physics but also fundamental physics in these highly magnetized accreting systems.

Figure 10: The frequency history from GBM pulsar monitoring (black diamonds) along with the simulated detections from the WFM (red disks). The frequencies were modeled from results by the Fermi/GBM and the source fluxes were taken from Swift/BAT. We used the source spectral model proposed by (Pradhan et al., 2014). An improved detection rate where the source dips below GBM's sensitivity is due to WFM's sensitivity below 8 keV. Compared to the GBM, pulsations are also detected earlier and with higher resolution by the WFM, enabling us to monitor the pulsed portion of the outburst during the critical onset stages of the event. The WFM will also be able to follow at any time the spectral evolution of the source in a wide energy band and with good resolution (as shown in the bottom panel).

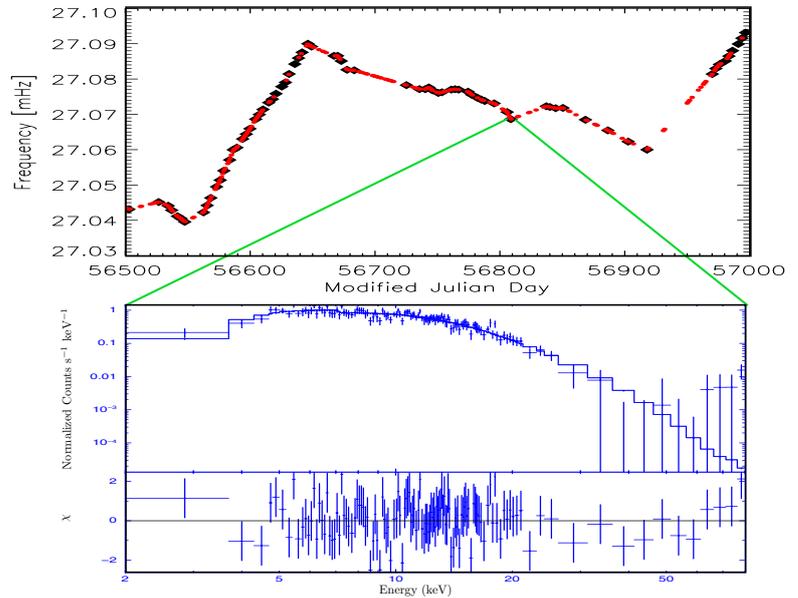

**The WFM is a perfectly suited instrument for these studies in that it is a wide field monitor capable of simultaneous measuring the source flux and accurately determining its spin frequency. Its lower energy threshold will make it the most sensitive monitor for these types of systems that has ever flown.** Measurements from the WFM of pulsed flux, un-pulsed flux, frequency and frequency derivatives of accreting X-ray pulsars will contribute to the solution of the questions mentioned above, and it will also trigger target of opportunity observations with the LAD at the onset of torque reversals. Early detections of these state changes is critical to initiate early observations by the LAD in order to understand the associated spectral/timing variations. To illustrate the WFM capabilities and compare them with those of existing facilities, we show in Fig. 10 the simulated spin frequency history of the win-fed HMXB OAO 1657-415 over 500 days and that derived from Fermi GBM measurements (see http://gammaray.msfc.nasa.gov/gbm/science/pulsars.html). The improved sensitivity of the WFM allows us to detect pulsations at lower luminosities, and virtually gap-free monitoring observations open up access to longer pulsation periods than any other instrument ever flown.

## 4.2 Phase resolved spectroscopy of superorbital modulations in HMXBs

Superorbital modulation, i.e. periodic modulation on a timescale longer than the orbital period, is well known in X-ray binaries such as Her X-1, SMC X-1, LMC X-4 and SS 433, where accretion occurs by Roche-lobe overflow and the driving mechanism is believed to be precession of the accretion disk, a jet, or the compact object itself. Irradiation of the accretion disk by the central X-ray source provides a possible mechanism for driving precession in NS systems (e.g. Ogilvie & Dubus, 2001, and references therein). In some cases the precessing disk is also believed to collimate relativistic jets, inducing the latter to precess as well. The jets in the Galactic microquasar SS 433 move with velocity close to $0.3c$, and multiple fluorescence lines originating in jets appear red or blue shifted, varying periodically during the $\sim 163$ d precession cycle. Being relatively bright in X-ray, optical, and radio domains, this source provides a unique test-bed for jet formation models. A monitoring of the X-ray line behavior (most notably the iron emission $K_\alpha$ line) provides a key tool for these studies, and can be easily carried out with the WFM. In Fig. 11 we present a simulated WFM spectrum for two precession phases of SS 433. The iron lines detected at high significance in the spectra provide an accurate measurement of the red and blue-shift caused by the two jets moving in opposite directions. **Similar observations can be obtained as a by-product of the long continuous sky monitoring that the WFM will be carrying out during the mission lifetime and will allow to improve the kinematic jet precession model**. Obviously, long term variations of the X-ray continuum can be studied in SS 433, as well as in many other relatively bright X-ray sources.

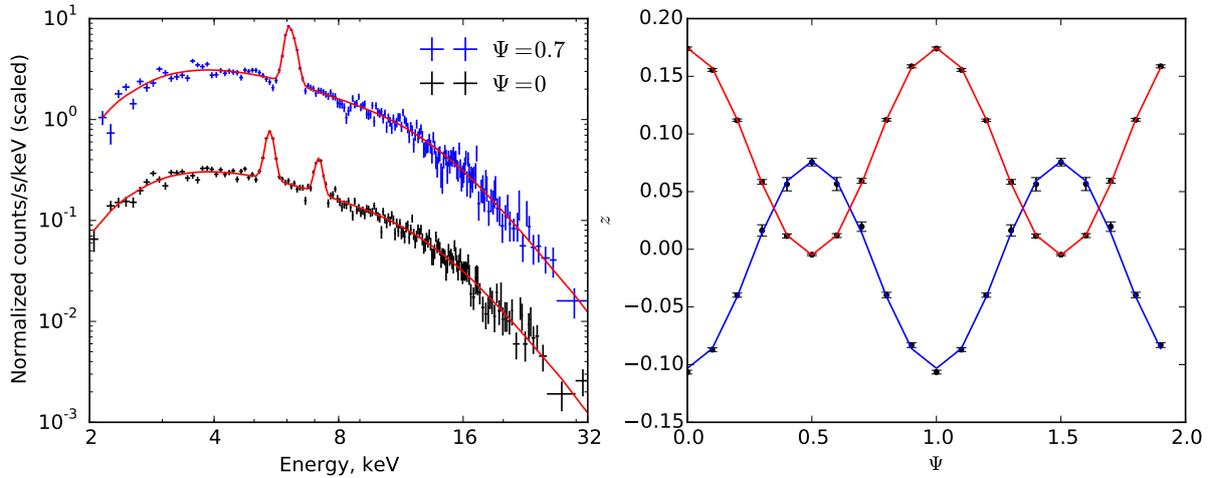

Figure 11: *Left:* simulated WFM spectra of the Galactic microquasar SS 433 for two superorbital phases (total integration time 3 Ms; note that large integration times are not an issue here, because we are studying the spectral variability of the source on a long time scale - the superorbital period in SS 433 is ~160 d). *Right:* reconstructed redshifts of the two iron lines (red and blue shifted) originating in the precessing jets of SS 433 as a function of the superorbital phase (Margon et al., 1984).

A different type of superorbital variability is also detected in several HMXBs, such as 2S 0114+650, IGR J16493-4348, 4U 1909+07, IGR J16418-4532, and IGR J16479-4514 (Corbet et al., 1999; Wen et al., 2006; Farrell et al., 2008; Corbet & Krimm, 2013). Corbet & Krimm (2013) found that for all five sources the ratio of superorbital period to orbital period is roughly a factor of 3. In addition, there is an apparent strong correlation between superorbital period and orbital period. Although the sample of HMXBs displaying superorbital modulation is growing, the nature of such modulation remains unknown. In particular, it remains to be clarified which is the mechanism driving superorbital modulation in sources where there is neither a precessing disk nor a jet, and if other classes of HMXBs exhibit a superorbital modulation as well. Thanks to the continuous monitoring that the WFM will perform on many HMXBs, all these open questions will be within reach for this instrument. *The WFM good sensitivity, energy resolution, and wide energy bands are expected to permit detailed spectroscopic investigation of all known and newly discovered superorbital modulations.*

## 5 Gamma-ray binaries with *LOFT*

γ-ray-loud binaries (GRLBs) are a class of massive X-ray binaries in which the interaction of an outflow from the compact object (black hole or NS) with the wind and radiation emitted by a hot, young star (either O or Be star) leads to the production of very high energy emissions (VHE; see Dubus, 2013, for a recent review). The prototype systems in this class are PSR B1259-63 (Aharonian et al., 2005), LS 5039 (Aharonian et al., 2006), LS I +61° 303 (Albert et al., 2006), H.E.S.S. J0632+057 (Aharonian et al., 2007), and 1FGL J1018-5859 (H.E.S.S. Collaboration et al., 2012; Fermi LAT Collaboration et al., 2012). Observations carried out with the Fermi/LAT telescope helped to reveal also different kinds of binaries emitting at VHE, as the microquasar Cyg X-3 (Fermi LAT Collaboration et al., 2009) and the symbiotic binary V 407 Cygni (Abdo et al., 2010).

Details of the physical mechanisms leading to the production of VHE emissions from GRLBs are still poorly understood, and it is possible that these systems are fundamentally different from the accretion-powered X-ray binaries. The nature of the compact object is known only for PSR B1259-63, where collision of the pulsar wind with the wind of the companion star leads to particle acceleration and γ-ray emission (Johnston et al., 1992). Other GRLBs might work similarly (the "hidden pulsar model"), but the pulsed radio emission from the NS can go undetected due to absorption in the companion's wind (due to the shorter orbital periods; Zdziarski et al., 2010). The orbital multi-wavelength variability pattern of a system like PSR B1259-63 is determined mostly by the details of the interaction of a relativistic pulsar wind with the strongly anisotropic wind of the companion (Johnston et al., 2005; Aharonian et al., 2005; Chernyakova et al., 2009, and references therein). The energy of the relativistic electrons in the pulsar wind is not known, and thus the observed X-ray emission can be explained either with synchrotron or inverse Compton (IC) mechanisms. Despite the recent intensive observational campaigns, it was still not possible to distinguish among these two possibilities (as well as investigate the composition of the pulsar wind; Chernyakova et al., 2009; Abdo et al., 2011). If the X-rays are due to IC mechanisms, the energies of the electrons responsible for such emission should be in the ∼10 MeV range. In the synchrotron model, the X-ray emitting electrons have multi-TeV energies. H.E.S.S. and Fermi are not sensitive enough to constrain the spectrum of PSR B1259-63 (Abdo et al., 2011), but with CTA it would be possible to better constrain its high energy part and find out the origin of the GeV and TeV energies. Simultaneous observations with the LAD will timely complete the energy coverage of the source spectrum and provide additional clues on the nature of its high energy emission by accurately measuring (within ∼1 ks) any possible spectral break, as suggested by Suzaku observations (Uchiyama et al., 2009). In the synchrotron model the break could indicate a transition from the synchrotron to the IC cooling in the Klein-Nishina regime. In the IC model, the break could result from either Coulomb/adiabatic cooling, or the presence of a low-energy cut-off in the electron spectrum (Chernyakova et al., 2009).

In a number of GRLBs, fast X-ray flares are also observed, and have been ascribed to the clumpy wind of the companion star (see, e.g., the case of LS I +61° 303; Zdziarski et al., 2010). Detailed studies of the broad-band spectral evolution of the source during these events could help to test the "hidden pulsar model", as the clump illumination by the pulsar wind will produce different patterns than those expected in the case of accretion (see, e.g., Romero et al., 2007, and references therein). *LOFT* observations will thus for the first time test the spectral variability of the system on the required short time scales. Simulations show that the LAD will be able to accurately measure spectral variations in, e.g., LS I +61° 303 within 1 ks, while hardness ratios can be measured with high SNR in time scales as short as 100 s.

While it is possible that all the classical GLRBs observed at TeV range contain a hidden pulsar, both AGILE and Fermi/LAT observations proved that also the classical microquasar Cyg X-3 is emitting up to GeV energies (Fermi LAT Collaboration et al., 2009; Piano et al., 2012). In this source, soft X-rays are thought to arise from an accretion disc surrounding the compact object, the hard X-rays from a hot corona above the disk, and the radio from the jet (Szostek et al., 2008). Paerels et al. (2000) showed that Cyg X-3 harbours a large number of emission lines. The iron line complex consists of helium-like and hydrogen-like iron ions at 6.7 keV and 6.9 keV, respectively, and cold iron K$\alpha$ line at 6.4 keV. The simultaneous usage of timing and spectral information is required to interpret the data unambiguously (see, e.g., Koljonen et al., 2013). As shown in Figure 12, the LAD will disentangle all the spectral components of the source within integration times <1 ks.

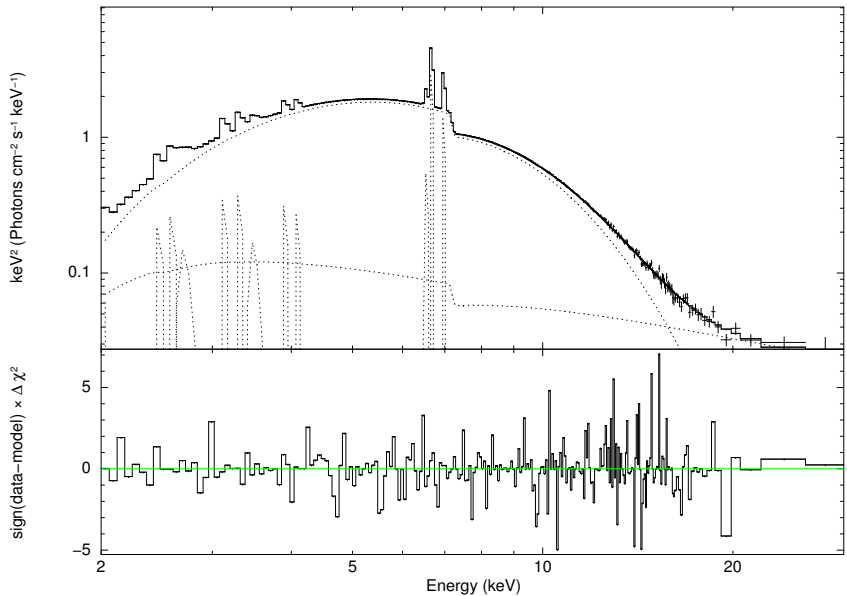

**Figure 12:** Simulated LAD spectrum of Cyg X-3 (the spectral model used includes the relevant emission lines, the contribution from the accretion disk and the jets; see Sect. 5). Thanks to its high throughput, the LAD will permit for the first time a study of the spectral-timing variability of the source on short time scales along the orbit, using the whole wealth of information coming from spectral lines and the simultaneous time series in a broad energy range. It will be possible to fully disentangle all the different spectral components of the source and see all of them in play simultaneously. In this example, an integration time of 1 ks was used.

## 6  *LOFT* and the ultra-luminous X-ray sources

ULXs are observationally defined as non-nuclear extragalactic X-ray point sources with luminosity exceeding the Eddington limit for a $\sim 10\,M_\odot$ compact object. The evidence gathered until now remains consistent with alternative interpretations that have important cosmological implications (e.g. Zampieri & Roberts 2009, Feng & Soria 2011 and references therein). Beside a massive primary star, ULXs could either contain an unprecedented class of intermediate mass black holes (IMBHs), that may represent the relics of the population of primordial BHs in the early Universe, or be BHs of stellar origin powered at extreme super-Eddington accretion rates, in physical conditions similar to those occurring in the first generation of Quasars at very high redshift. Recent dynamical measurements or constraints to the BH mass have been derived and are in agreement with the latter interpretation (Liu et al. 2013; Motch et al. 2014). While observationally IMBHs of several hundreds to thousands $M_\odot$ are not required for the majority of ULXs (e.g. Zampieri & Roberts 2009), a handful of hyperluminous ($> 10^{41}$ erg s$^{-1}$, HLXs; Farrell et al. 2009) or very luminous ($> 5 \times 10^{40}$ erg s$^{-1}$; Sutton et al. 2012) sources have been identified that are good candidates for harbouring them (in one case also thanks to its timing properties; e.g. Pasham et al. 2014). On the other hand, some transient ULXs may even host NSs (Bachetti et al. 2014).

So far, X-ray observations have shown that the spectra of ULXs are characterized by properties not commonly observed in Galactic BH X-ray binary systems accreting at sub-Eddington rates (Stobbart et al. 2006; Gonçalves & Soria 2006). A roll-over at high energy, usually at 3-5 keV, is often observed coupled to a soft excess (Stobbart et al. 2006). Gladstone et al. (2009) found that such a curvature is almost ubiquitous in the highest quality ULX spectra and proposed that it is a characteristic feature of a new spectral state, the *Ultraluminous state*. A few bright ULXs very recently observed with NuSTAR show that the high energy curvature persists also above 10 keV (Bachetti et al. 2013; Walton et al. 2014). Complex long-term X-ray spectral variability is observed, often different from source to source, in some cases understandable within the framework of an optically thick, comptonizing corona/wind energetically coupled to and possibly partially obscuring an accretion disc (Middleton et al. 2011b; Sutton et al. 2013; Pintore et al. 2014).

**Figure 13:** Simulated LAD PSD for a 20 ks LAD observation of M 82 X-1, assuming a source count rate of 180 cts/s (~0.8 mCrab) and a background count rate of ~2500 cts/s (including the instrumental sky and host galaxy background). The binning time of the light curve is 0.1 s. The input parameters for the QPO and the band-limited noise are taken from Mucciarelli et al. (2006). The simulated PSD shows a clearly detected QPO with a rms of 1% in the 0.01-5 Hz frequency range and a very large significance (~27σ). For comparison, we also show the PSD fit of a 20 ks LAD observation assuming a source count rate of 140 cts/s (~0.5 mCrab, significance ~21σ) and that of the 66 ks *XMM-Newton* observation reported by Mucciarelli et al. (~0.5 mCrab, significance ~9σ; 2006).

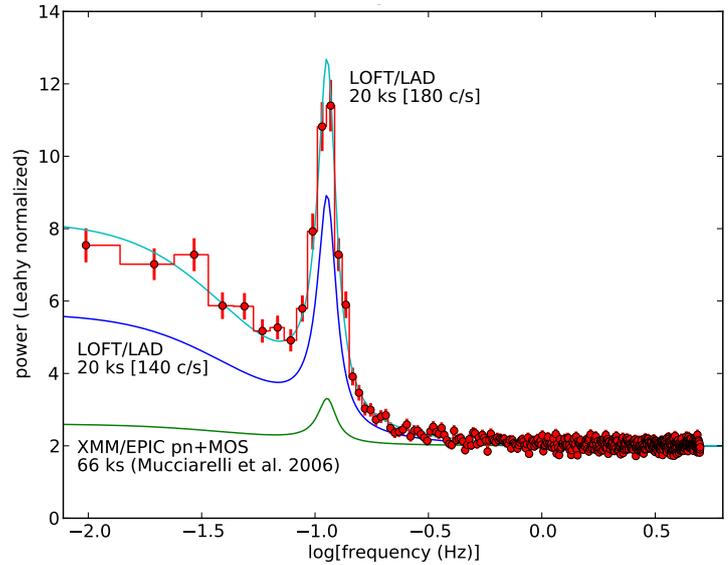

The X-ray timing properties of ULXs are also poorly understood. Heil et al. (2009) analysed the PSD of a sample of 16 bright ULXs and found two groups of sources: the first displays variability at about the same level (~ 10%), while in the second the variability is almost absent. A handful of ULXs (M82 X-1, Strohmayer & Mushotzky 2003; X42.3+59 in M82, Feng et al. 2010; NGC 5408 X-1, Strohmayer et al. 2007) show quasi periodic oscillations (QPOs) with centroid frequencies ranging from a few mHz (X42.3+59) up to ~170 mHz (M82 X-1), but their classification is unclear (e.g. Middleton et al. 2011a; Caballero-García et al. 2013). In M82 X-1 twin-peak QPOs above 1 Hz have been identified in the PSD, suggesting a BH mass of ~ 400 $M_\odot$ (Pasham et al. 2014). The recent identification of a periodicity of 1.37 s in the transient ULX X42.3+59 in M82 by NuSTAR (Bachetti et al. 2014) opens up a new discovery space at hard X-ray energies, that may reveal a hitherto uncovered population of ULXs containing NSs accreting largely above Eddington.

A ULX with a luminosity ≥4×10$^{40}$ erg s$^{-1}$ above 2 keV and located at distances smaller than ~4 Mpc has an observed flux ≥0.8 mCrab (2×10$^{-11}$ erg cm$^{-2}$ s$^{-1}$) and then sufficient counts on the LAD to measure strong features in the PSD with high significance in a reasonable exposure time. M82 X-1 is often observed at this flux level and its QPOs below 200 mHz would then be easily detectable by the LAD. We simulated low frequency QPOs with properties similar to those in M82 X-1 and found that they are detectable in 20 ks with a high significance (Figure 13). These simulations take into account the additional galaxy background count rate produced by the diffuse emission and integrated point-like population of sources in M82 (estimated using the spectral model of Strickland & Heckman 2007). Strong M82 X-1-like low-frequency QPOs in less luminous sources are also detectable. A 20 ks exposure gives a detection significance of ~7σ at a flux level of ~0.3 mCrab, while 100 ks are needed for a ~3σ detection at a flux level of ~0.15 mCrab. The (observed) flux distribution of ULXs (Walton et al. 2011; Pintore et al., in preparation) shows that 6 sources have a maximum observed flux (0.3-10 keV) above ~0.3 mCrab and 8 sources above ~0.15 mCrab. The majority of them are known to undergo significant short-term variability in the *XMM-Newton* bandpass and are then very promising targets for *LOFT*.

Twin-peak QPOs above 1 Hz, with properties similar to those reported by Pasham et al. (2014), can also be detected at the ~4σ level with a 350 ks exposure. Prospects for the LAD identification of periodic signals in bright transient and nearby ULXs are also good. A sinusoidal signal with properties similar to that discovered in X42.3+59 in M82 (frequency ~1 Hz, pulsed fraction ~15%; Bachetti et al. 2014) can be easily detected at flux levels ≥3.5×10$^{-12}$ erg cm$^{-2}$ s$^{-1}$ (≥6σ) in a ≤120 ks exposure.

The *LOFT* WFM will potentially discover very luminous transient ULXs or HLXs. Assuming a goal sensitivity of 2 mCrab in 50 ks, the WFM could detect a transient with $L_{>2\,\mathrm{keV}} \approx 10^{41}$ erg s$^{-1}$ up to a distance of ≈4 Mpc.

Recent analysis of the statistical properties of massive BH binaries dynamically interacting in clusters shows that the fraction of BHs which undergoes unstable accretion can be up to 2 times more numerous than that undergoing stable Roche Lobe overflow (Mapelli et al. 2013). Assuming a duty cycle for the transient systems of about 20% (Burke, M. J., 2013, PhD Thesis, The University of Birmingham), the recurrence time may be as short as few years (from the shortest observed outburst duration, e.g. M 32 ULX-2, Esposito et al. 2013). From the ratio of unstable/stable systems and the assumed duty cycle, under favourable assumptions we would then expect to detect 1 transient every 2 persistent ULXs. If the flux distribution of transient sources is similar to that of persistent sources, at least one bright transient should be detectable by WFM above ~2 mCrab every few years. The potential confusion with other very bright X-ray transient events (e.g. TDEs or X-ray bright SNe) should not be of major concern as the X-ray spectrum and light curve are significantly different.

## 7   Conclusion

*LOFT*'s unprecedented combination of large effective area, high time resolution, and good spectral resolution for the LAD will for the first time enable studies of stellar winds, accretion flows, and accretion columns on dynamical and sub-dynamical timescales in HXMBs, overcoming long-standing difficulties caused by having to average data over long timescales in disentangling physical properties. *LOFT*'s WFM will be the most sensitive wide-field monitor ever flown in the 2-50 keV band, enabling studies of HMXB outburst onsets, flares, accretion torques, and new discoveries. Especially of interest now that NuSTAR has discovered a pulsar in a ULX system (Bachetti et al., 2014), the LAD will provide unparalleled studies of ULX timing properties and the WFM will provide unique opportunities to discover new ULXs. If flown as the M4 mission, LOFT provides unique potential as an X-ray observatory in a wide variety of astrophysical studies that go far beyond the core science.